\documentclass[preprint,3p]{article}

\usepackage[utf8]{inputenc}
\usepackage{amssymb}
\usepackage{bm}
\usepackage{enumerate}
\usepackage{booktabs}
\usepackage{graphicx}
\usepackage{color}
\usepackage{subfigure}
\usepackage{caption}
\usepackage{yhmath}
\usepackage{amsmath}
\usepackage{url}

\newcommand{\mE}{\mathcal{E}}
\newcommand{\mG}{\mathcal{G}}
\newcommand{\mF}{\mathcal{F}}
\newcommand{\mH}{\mathcal{H}}

\newcommand{\mT}{\mathcal{T}}

\newcommand{\mV}{\mathcal{V}}

\newcommand{\Sij}{\mathcal{S}_{ij}}
\newcommand{\Skk}{\mathcal{S}_{kk}}

\newcommand{\bu}{\mathbf{u}}

\newcommand{\bF}{\mathbf{F}}

\newcommand{\bmG}{\boldsymbol{\mG}}

\newcommand{\bS}{\mathbf{S}}

\newcommand{\bU}{\mathbf{U}}

\newcommand{\bq}{\mathbf{q}}

\newcommand{\wt}{\widetilde}

\newcommand{\fsigmaij}{\wt{\sigma}_{ij}}

\newcommand{\de}{\partial}
\newcommand{\Div}{\nabla\cdot}

\newcommand{\w}{\omega}

\newcommand{\sigmaij}{\sigma_{ij}}

\newcommand{\tauij}{\tau_{ij}}
\newcommand{\ei}{e_{\rm i}}

\newcommand{\bn}{\mathbf{n}}
\newcommand{\bx}{\mathbf{x}}
\newcommand{\br}{\mathbf{r}}

\newcommand{\RE}{Re}
\newcommand{\MA}{M\hspace{-1pt}a}
\newcommand{\PR}{Pr}

\newcommand{\les}{\rangle_{\scriptscriptstyle{\mathcal{F}}}} 
\newcommand{\rans}{\rangle_{\scriptstyle{\mathcal{E}}}} 
\newcommand{\hyb}{\rangle_{\scriptstyle{\mathcal{H}}}}
\newcommand{\lrho}{\langle \rho}
\newcommand{\UTj}{\langle \widetilde{u}_j\hyb}
\newcommand{\UTi}{\langle \widetilde{u}_i\hyb}
\newcommand{\RH}{\lrho \hyb}
\newcommand{\ET}{\langle \widetilde{e} \hyb}

\begin{document}
\begin{center}
\Large{\textbf{A DG Implementation of a Novel Hybrid RANS/LES Technique With RANS Reconstruction}}
\end{center}
\vspace{2em}

\begin{center}
{{\large Antonella Abbà$^1$, Massimo Germano$^2$, Michele  Nini$^{1,*}$\let\thefootnote\relax\footnote{$^*$ \tt{michele.nini@polimi.it}}, Marco Restelli$^3$}}
\end{center}

\vspace{1em}
\begin{center}
{$^1$ Department of Aerospace Science and Technology, Politecnico di Milano\newline Via La Masa, 34, 20156 Milano, Italy \\[4mm]
$^2$Department of Civil and Environmental Engineering, Duke University, Durham, North Carolina 27708, USA\\[4mm]
$^3$ NMPP -- Numerische Methoden in der Plasmaphysik Max--Planck--Institut f\"ur Plasmaphysik, Boltzmannstra\ss e 2, D-85748 Garching, Germany}
\end{center}

\vspace{1em}


%

\begin{abstract}
\noindent
A new hybrid RANS/LES technique, based on the hybrid filter proposed by Germano ~\cite{germano:2004}, has been studied. The novelty herein introduced is represented by the reconstruction of the Reynolds stress tensor. As a consequence, no explicit RANS model is needed. The RANS and LES terms are merged using a constant blending factor.

\noindent
The model is implemented in a numerical code based on a high order Discontinuous Galerkin (DG) finite element formulation.

\noindent
The test case considered for numerical simulations is the turbulent turbulent channel flow at Mach = $0.2$. The comparison with available DNS data shows a good agreement and, in general, an improvement with respect to pure LES results, confirming that the technique herein proposed represents a promising approach to the numerical simulation of turbulent flows.

\vspace{1em}

\end{abstract}

\section{Introduction}

For a wide range of applications the grid resolution required by a \emph{Large Eddy Simulation} (LES) is still too expensive, especially in the wall bounded flows where the size of turbulent structures requires a resolution similar to the ones required by \emph{Direct Numerical Simulation} (DNS). On the other hand, the cheaper \emph{Reynolds Averaged Navier--Stokes} (RANS) methods do not provide the amount of information required in many simulations. Therefore, combining LES approach with RANS  models represents a reasonable choice to obtain an appropriate description of turbulent flows with a feasible cost.
As a consequence, since in 1997 Spalart \cite{spalart:1997} proposed the \emph{Detached Eddy Simulation}, hybrid RANS/LES methods have become a very interesting topic in aerodynamics research.

In the last years, several hybrid methods have been proposed. An accurate review and classification can be found in the work presented by Fr\"olich and Von Terzi in 2008\cite{frolich:2008}.
The authors define three different categories of hybrid methods: \emph{unified} models which exploit the structural similarity of RANS and LES, using the same transport equations for both techniques and then obtaining the transition changing the model coefficients;  \emph{segregated} models characterized by two different domains for RANS and LES with, in general, a discontinuous solution at the interface between the two regions  and, finally, the  \emph{second generation U-RANS} based on unsteady RANS model, without grid dependencies and, usually, with damping factors related to the amount of resolved turbulent fluctuations.

Hybrid RANS/LES techniques have obtained a good success and, some of them (e.g. DDES \cite{spalart:2006}), have reached a high popularity and are often used both in research and industrial applications.
Nevertheless, they still present some critical aspects, in particular at RANS/LES interfaces. In this region there are problems in terms of momentum and energy transfer, leading to incorrect predictions in velocity profiles and skin friction. The most common strategy to overcome these obstacles is adding a stochastic forcing term \cite{piomelli:2003} or, similarly, using a \emph{back--scatter} model \cite{davidson:2009}.
 
A different strategy is represented by the hybrid filter methods. In these methods the equations are  derived applying the hybrid filter directly to the Navier-Stokes equations. A specific formulation of this family, which we  consider in this work, is the additive filter proposed by Germano in 2004 \cite{germano:2004}. Another approach is represented by the spatial filter proposed by Hamba in 2011 \cite{hamba:2011}. Germano's additive filter is obtained combining a statistical operator RANS with an LES filter. Applying this hybrid filter to NS equations we obtain exact equations which contain RANS and LES terms. Moreover, these equations already include terms which are capable of representing the interactions between RANS and LES. Therefore, no artificial forcing terms are needed.

Germano's hybrid filter approach has already been studied by Rajamani and Kim \cite{rajamani:2010} who have performed both \emph{a priori} and \emph{a posteriori} tests for incompressible case, and by Sanchez-Rocha and Menon\cite{sanchez:2009} \cite{sanchez:2011} who have derived and approximated equations for compressible flow. 

The main novelty herein introduced is represented by the treatment of RANS terms, which are reconstructed using hybrid and LES stresses and resolved velocity field. 

Numerical simulations have been conducted using the Variational Multiscale (VMS) framework \cite{hughes:1998} combined to Discontinuous Galerkin (DG) finite element method. Thanks to the possibility of using arbitrary meshes, its good parallel scalability and its accuracy, this numerical approach seems very suitable for CFD calculation.
 Moreover,in the case of LES simulations, a further advantage is the possibility of defining a filter simply by projecting the solution on a lower oreder polynomial space. This turns to be very useful for the dynamic procedure \cite{germano:1991}, in which different filtering levels are required.

Examples of DG applications to fluid dynamics can be found in \cite{bassi:2005} for RANS, while for a DG-VMS approach we remand to \cite{collis:2002}, \cite{vanderbos:2007} and \cite{abba:2014}.

In section 2 the compressible Navier--Stokes equations are presented and filtered. The corresponding hybrid model terms and RANS reconstruction process are described in section 3, while the numerical methodis presented in section 4. Finally, the numerical results are shown in section 5. 

\section{Mathematical formulation}
We start from the compressible Navier--Stokes equations in dimensionless
form:
\begin{subequations}
\label{eq:nscompr}
\begin{align}
&\de_t \rho + \de_j (\rho u_j) = 0 \\
&\de_t (\rho u_i) + \de_j (\rho u_i u_j) + 
\frac{1}{\gamma\,\MA^2}\de_i p - \frac{1}{\RE}\de_j \sigmaij =
0 \label{eq:nscompr-momentum} \\
&\de_t (\rho e) + \de_j (\rho h u_j)
- \frac{\gamma\,\MA^2}{\RE} \de_j (u_i \sigmaij)  + \frac{1}{\kappa\RE\PR}\de_j q_j = 0 ,
\end{align}
\end{subequations}

\noindent
where $\rho$, $\bu$ and $e$ denote dimensionless density, velocity and
specific total energy, respectively, $p$ is the pressure,
 $h$ is the specific enthalpy defined by $\rho h=\rho e+p$, and
$\sigma$ and $\bq$ are the diffusive momentum and heat fluxes.
$\gamma=c_p/c_v$ ($1.4$ in the case studied) is the ratio between the specific heats at constant pressure  and volume respectively.
The Mach number $\MA$, the Reynolds number $\RE$ and the Prandtl number $\PR$
are defined as 
\begin{equation}
\MA = \frac{V_r}{\left( \gamma R T_r \right)^{1/2}}, \qquad
\RE = \frac{\rho_rV_rL_r}{\mu_r}, \qquad \PR = \frac{c_p }{\kappa}
\label{eq:adim-numbers}
\end{equation}
on the basis of appropriate reference quantities (denoted with $r$), of the
ideal gas constant $R=c_p - c_v$ and $\kappa=R/c_p$.

In order to complete the system (\ref{eq:nscompr}) we also need the state equation for an ideal gas in dimentionless form, given by
\begin{equation}
p = \rho T,
\label{eq:state-eq}
\end{equation}

The temperature $T$ is related to the energy equation by means of the specific internal energy $\ei$
\begin{equation}
e = \ei + \frac{\gamma\MA^2}{2} u_ku_k,
\qquad \ei = \frac{1-\kappa}{\kappa} T.
\label{eq:ei}
\end{equation}
 Finally, the model is closed with the constitutive equations for the diffusive fluxes
\begin{equation}
\sigmaij = \mu \Sij^d, \qquad
q_i = -\mu\de_i T,
\label{eq:constitutive}
\end{equation}
with $\Sij = \de_j u_i + \de_i u_j$ and $\Sij^d = \Sij -
\dfrac{1}{3}\Skk\delta_{ij}$. 
The dynamic viscosity $\mu$ is assumed to depend only on
temperature $T$ in agreement with Sutherland's hypothesis (see e.g.
\cite{schlichting:1979}).

The hybrid equations are obtained applying the hybrid  filter to system \ref{eq:nscompr}.

Germano's hybrid filter is defined as:
\begin{equation}
\label{eq:h-filter}
\mH= k \mathcal{F} + (1-k) \mE,
\end{equation}
where $\mathcal{F}$ and $\mathcal {E}$ represent the LES filter and the statistical operator (i.e. RANS operator). $k$ is a blending factor which can vary between $1$, resulting in a pure LES,  to $0$ yielding a pure RANS.

Coherently with \cite{germano:2004} we assume that:
\begin{equation}
\label{assumption}
\mE\mH= \mE ,\qquad \mE\mathcal{F}= \mE, \qquad \mathcal{F}\frac{\partial}{\partial x} = \frac{\partial}{\partial x}\mathcal{F}. 
\end{equation}
Notice that the last assumption in (\ref{assumption}) is the standard assumption of commutativity between filtering and differentiation for LES models. Such an assumptions is not satisfied by the operator F considered here. However, we will ignore the resulting error, as it is often done in LES modelling, when a non-uniform filter is used   \cite{vanderbos:2005}. Considering the hybrid filter, we observe that it does not commute with space and time derivative. In fact, we have:

\begin{equation}
\label{eq:com}
 \mH\frac{\partial}{\partial x} = \frac{\partial}{\partial x}\mH - \frac{\partial}{\partial x} k \left( \mathcal{F} -\mE \right).
\end{equation}
Although, in general, for hybrid methods we want to move from a pure RANS near the wall to a pure LES in the freestream region, in this preliminary study a constant blending factor is considered. Hence, all the terms related to the non commutativity vanish.

Moreover, in order to avoid additional subgrid terms, we also introduce a
Favre-average, defined for a generic variable $\psi$ as
\begin{equation}
\label{eq:favre}
\langle \widetilde{\psi} \hyb = \frac{\lrho \psi \hyb}{\lrho \hyb}.
\end{equation}
Finally, applying (\ref{eq:h-filter}) and (\ref{eq:favre}) to (\ref{eq:nscompr}),we obtain:

\begin{subequations}
\label{eq:fnscompr}
\begin{align}
&\de_t \RH + \de_j (\RH \UTj) = 0 \\
&\de_t (\RH \UTi) + \de_j (\RH \UTi \UTj) + 
\frac{1}{\gamma\,\MA^2}\de_i \langle p\hyb -\\
& \hspace{6cm} \frac{1}{\RE}\de_j \langle \widetilde {\sigmaij} \hyb = - \de_j \tauij^{\mH} \left(\rho,u_i, u_j\right) \label{eq:fnscompr-momentum} \\
&\de_t (\RH \ET) + \de_j (\RH \langle h\hyb \UTj)
- \frac{\gamma\,\MA^2}{\RE} \de_j (\UTi \langle \widetilde{\sigmaij}\hyb) + \\
& \hspace{6cm}\frac{1}{\kappa\RE\PR}\de_j \langle \widetilde{q_j} \hyb =    - \de_j  \vartheta_j^{\mH}\left(\rho,h,u_j\right) .
\end{align}
\end{subequations}
 In the derivation of (~\ref{eq:fnscompr}), we have considered the following assumptions, which are consistent with  \cite{piomelli:2000} and \cite{vreman:1995}: 
\begin{equation}
 \langle{\sigma}_{ij}\hyb \approx \langle\fsigmaij\hyb, \qquad
 \langle{q}_i\hyb \approx \langle\wt{q}_i\hyb, \qquad
 \langle u_j \sigmaij\hyb \approx \UTj\langle \widetilde{\sigmaij}\hyb. \notag
 \end{equation}

Therefore, the only sub-grid terms to be modelled are 
\begin{subequations}
\begin{align}
\tauij^{\mH} \left(\rho,u_i, u_j\right) =& \lrho u_i u_j \hyb -\RH \UTi \UTj \label{eq:m_sgs} \\
\vartheta_j^{\mH}\left(\rho,h,u_j\right)=& \lrho h u_j \hyb - \RH \langle \widetilde{h} \hyb \UTj.  \label{eq:en_sgs}
\end{align}
\end{subequations}

\section{Model description}

\subsection{Momentum equation}
Here we will consider only a nearly incompressible flow, so that the \emph{sub-grid stress tensor} (\ref{eq:m_sgs})  can be approximated as

\begin{equation}
 \tauij^{\mH} \left(\rho,u_i, u_j\right) \approx \RH  \tauij^{\mH} \left(u_i, u_j\right).
\end{equation}
\noindent
Therefore, using the definition for the generalized central moment of second order \cite{germano:1992}, we arrive at

\begin{align} 
\label{eq:tau_h}
 \tau^{\mH}(u_{i},u_{j}) = & k \tau^{\mathcal{F}}(u_{i},u_{j}) + (1 - k) \tau^{\mE}(u_{i},u_{j}) +  \notag \\ & k(1-k)( \langle u_{i}\rangle _{\mathcal{F}} - \langle u_{i}\rangle _{\mE} )( \langle u_{j}\rangle _{\mathcal{F}} - \langle u_{j}\rangle _{\mE} ),
\end{align}

where 
$\tau^{\mathcal{F}}(u_{i},u_{j})$ is the LES term, $\tau^{\mE}(u_{i},u_{j})$ is the RANS term and $k(1-k)( \langle u_{i}\rangle _{\mathcal{F}} - \langle u_{i}\rangle _{\mE} )( \langle u_{j}\rangle _{\mathcal{F}} - \langle u_{j}\rangle _{\mE} )$ represents the \emph{Germano} stress ~\cite{rajamani:2010}.

The filtered velocity  $\langle u \les$ can be obtained from

\begin{equation}
\label{u_f}
\langle u_{i} \rangle ^{\mathcal{F}} = \frac{\langle u_{i}\rangle ^{\mH} - (1 - k)\langle u_{i}\rangle ^{\mE} }{k},
\end{equation}
and using (\ref{u_f}) we have
\begin{align} 
\label{eq:tau_h_mod}
 \tau^{\mH}(u_{i},u_{j}) = & k \tau^{\mathcal{F}}(u_{i},u_{j}) + (1 - k) \tau^{\mE}(u_{i},u_{j}) +  \notag \\ & \frac{1-k}{k}( \langle u_{i}\rangle _{\mH} - \langle u_{i}\rangle _{\mE} )( \langle u_{j}\rangle _{\mH} - \langle u_{j}\rangle _{\mE} ).
\end{align}
It is worth noting that (\ref{eq:tau_h_mod}) can be closed by means of two arbitrary RANS and LES models. Concerning the LES model, in this work we have used an anisotropic dynamic model \cite{abba:2003}. Concerning the RANS field, it can be obtained either from previous DNS computations, or  from experimental results, or implicitly reconstructed from the hybrid and LES stress tensors and from the velocity field. The latter is the approach herein studied and tested, as we discuss in the next paragraph.

\subsection{RANS reconstruction}
 
\noindent
$\tau^{\mE}(u_{i},u_{j})$ can be written as
\begin{align} 
\label{eq:tau_e}
\tau^{\mE}(u_{i},u_{j}) = & \langle u_{i}u_{j} \rans -
                          \langle u_{i} \rangle \langle u_{j} \rangle = \notag \\[2mm]
                         = &\langle \langle u_{i}u_{j} \hyb \rans - \langle \langle u_{i}\hyb \rans \langle \langle u_{j}\hyb \rans  +  \\[2mm] 
                         & \langle \langle u_{i}\hyb \langle u_{j}\hyb \rans - \langle \langle u_{i}\hyb \langle u_{j}\hyb \rans = \notag \\[2mm]
                      = & \langle \tau_{H}(u_{i},u_{j}) \rans + \tau_{E} ( \langle u_{i}\hyb,  \langle u_{j}\hyb) \notag
\end{align}
where, splitting velocity at $\mH$ level in average and fluctuating part, $ \langle \mathbf{u} \rangle _{\mH} = \langle \langle \mathbf{u} \rangle _{\mH} \rangle _{\mE} +  \langle \mathbf{u} \rangle _{\mH} '$, the latter term becomes:

\begin{align}
\label{eq:
tau_eh}
\tau_{\mE} ( \langle u_{i}\rangle _{\mH},  \langle  u_{j}\rangle  _{\mH})  =\langle (\langle \langle u_{i} \rangle _{\mH} \rangle _{\mE}& +  \langle u_{i} \rangle _{\mH} ')(\langle \langle u_{i} \rangle _{\mH} \rangle _{\mE} +  \langle u_{i} \rangle _{\mH} ') \rangle _{\mE} - \langle \langle u_{i}\rangle _{\mH} \rangle _{\mE} \langle \langle u_{j}\rangle _{\mH} \rangle _{\mE} = \\[2mm]
& = \langle ( \langle u_{i}\rangle _{\mH} - \langle u_{i}\rangle _{\mE} )( \langle u_{j}\rangle _{\mH} - \langle u_{j}\rangle _{\mE} ) \rangle _{\mE}. \notag
\end{align}
\noindent
Substituting the hybrid stress tensor definition (\ref{eq:tau_h_mod}) in (\ref{eq:tau_e}) one obtains

\begin{align} 
\label{eq:tau_e_rec1}
\tau^{\mE}(u_{i},u_{j}) & = k \langle \tau^{\mathcal{F}}(u_{i},u_{j}) \rangle_{\mE} + (1 - k) \tau^{\mE}(u_{i},u_{j}) +  \\[2mm] &  \frac{1-k}{k} \langle ( \langle u_{i}\rangle _{\mH} - \langle u_{i}\rangle _{\mE} )( \langle u_{j}\rangle _{\mH} - \langle u_{j}\rangle _{\mE} ) \rangle _{\mE} + \notag \\[2mm]
 & ~ \tau^{\mE} ( \langle u_{i}\rangle _{\mH},  \langle u_{j}\rangle _{\mH}). \notag
\end{align}
Using now relation \ref{eq:tau_e}, the Reynolds stress tensor becomes:
\begin{equation} 
\label{eq:tau_e_rec2}
\tau^{\mE}(u_{i},u_{j}) = \langle \tau^{\mathcal{F}}(u_{i},u_{j}) \rangle_{\mE} + \frac{1}{k^{2}}\tau^{\mE} ( \langle u_{i}\rangle _{\mH},  \langle u_{j}\rangle _{\mH}) 
\end{equation}
\noindent
Inserting relation (\ref{eq:tau_e_rec2}) in (\ref{eq:tau_h_mod}), we can finally obtain the expression of $\tau^{\mH}(u_{i},u_{j})$, namely

\begin{align} 
\label{eq:tau_h_final}
 \tau^{\mH}(u_{i},u_{j}) = &  k \tau^{\mathcal{F}}(u_{i},u_{j}) +  \notag \\
 &(1 - k) \langle \tau^{\mathcal{F}}(u_{i},u_{j}) \rangle_{\mE} + \frac{1 - k}{k^{2}}\tau^{\mE} ( \langle u_{i}\rangle _{\mH},  \langle u_{j}\rangle _{\mH}) + \notag \\
 &  \frac{1-k}{k}( \langle u_{i}\rangle _{\mH} - \langle u_{i}\rangle _{\mE} )( \langle u_{j}\rangle _{\mH} - \langle u_{j}\rangle _{\mE} ).
\end{align}
A drawback of this procedure is represented by the presence of term $\frac{1}{k^2}$ in (\ref{eq:tau_e_rec2}), which leads to an ill conditioned problem for low values of $k$. In fact, although a lower limit for $k$ must be setted also in traditional approach, the square terms $k^2$ at the denominator leads to a greater value for this limit.

\subsection{Energy equation}

As shown in \cite{sanchez:2009,sanchez:2011}, the application of the  hybrid filter to energy equation leads to several additional terms, making modelling very costly and difficult. To avoid this problem, here a different approach has been adopted.

Following the guidelines given by Lenormand \cite{lenormand:2000} and Knight \cite{knight:1998} for the LES approximation of energy equation, the sub-grid stress tensor can be reduced to two contributions: heat flux ($Q$) and turbulent diffusion($J$).

Extending these assumptions to the dynamic--anisotropic model,  we have 
\begin{equation}
\label{eq:energy}
\vartheta_j= Q_j + J_j \approx \bar{\rho}\Delta^{2}|\mathit{S}|C^{Q}_{j} \partial _{j} T + \bar{\rho}\Delta^{2}|\mathit{S}|C^{J}_{j}  u_{k}\partial _{j} u_{k} - \tau_{jk}^{\mF} u_{k} - \frac{1}{2} u_{j} \tau_{kk}^{\mF},
\end{equation}
where $\mathit{S}$ represents the rate of strain tensor, and coefficient $C^{q}$  and $C^{J}$ are computed using a dynamic procedure.
 
In the proposed hybrid formulation proposed, the first two terms are the same of LES, while in the latter ones the $\tau^{\mF}$ is substituted by $\tau^{\mH}$, the same calculated for momentum balance by means of (\ref{eq:tau_h_final}). 

Thanks to this correction, hybrid terms enter into the energy equation modifying the turbulent diffusion. Considering the simplicity of the implementation and that it does not require any computational overhead, this seems to be a good compromise, especially at the low Mach number.

Notice that the resulting method turns to be rather general; in fact, it can be extended to any LES model in which sub-grid turbulent diffusion is modelled starting from the Knight proposal
\begin{equation}
J_j \approx \tau_{jk} u_{k} - \frac{1}{2} u_{j} \tau_{kk}
\end{equation}

\section{Numerical method}
The hybrid filtered Navier-Stokes equations presented in the previous sections are spatially discretized using the discontinuous Galerkin finite elements method. The approach herein employed is the same used in \cite{abba:2014} and follows the guidelines given by \cite{giraldo:2008} and more in general of the \emph{Local Discontinuous Galerkin} methods \cite{bassi:1997}. In this section a brief description of the discretization process is reported, for the details we refer to \cite{maggioni:2012}. 

In this framework Eq. ~\ref{eq:fnscompr} can be written as:

\begin{equation}\label{eq:csv_auxvar}
\begin{array}{l}
\displaystyle
\de_t \bU + \Div \bF^{{\rm c}}(\bU) - \Div \bF^{{\rm v}}(\bU,\bmG) 
 + \Div \bF^{{\rm sgs}}(\bU,\bmG) = \bS \\
 \displaystyle
 \bmG - \nabla{\boldsymbol \varphi} = 0,
\end{array}
\end{equation}
where $U=[\lrho\hyb, \lrho\hyb  \langle \wt{\mathbf{u}}\hyb$ and $\lrho\hyb \ET]^{T}$,$\boldsymbol \varphi=\left[ \langle \wt{\mathbf{u}}\hyb, \langle \wt{T} \hyb \right]^{T}$ collects the variables, whose gradients are required for flux computations, i.e. velocities and temperature.

The fluxes $\bF^{{\rm c}}, \bF^{{\rm v}}, \bF^{{\rm sgs}}$, respectively \emph{convective}, \emph{viscous} and \emph{sub-grid}, are given by

\[
\bF^{{\rm c}} = \left[
\begin{array}{c}
 \lrho\hyb\langle\wt{\bu}\hyb \\
 \lrho\hyb\langle\wt{\bu}\hyb\otimes\langle\wt{\bu}\hyb +
 \frac{1}{\gamma\MA^2}\langle p \hyb \mathcal{I} \\
 \lrho\hyb \langle\wt{h}\hyb \langle\wt{\bu}\hyb
\end{array}
\right],\quad
\bF^{{\rm sgs}} = \left[
\begin{array}{c}
 0 \\
 \tau^{\mH} \\
 \vartheta^{\mH}
\end{array}
\right]\]
\\
\[
\bF^{{\rm v}} = \left[
\begin{array}{c}
 0 \\
 \frac{1}{\RE}\langle\wt{\sigma}\hyb \\
 \frac{\gamma\MA^2}{\RE} \langle \wt{\bu}^T \hyb \langle \wt{\sigma}\hyb
 - \frac{1}{\kappa\RE\PR} \langle \wt{\bq}\hyb
\end{array}
\right]\]
where $\tau^{\mH}$ and $\vartheta^{\mH}$ are obtained from (\ref{eq:tau_h_final}) and (\ref{eq:energy}). We remark that this structure is absolutely general and is the same for LES, hybrid RANS/LES methods and also for \emph{unsteady} RANS. Therefore, according to the concept of \emph{implicit} filtering \cite{mason:1986}, we can choose the set of equation to be solved simply working on the sub-grid terms $\tau$ and $\vartheta$.

Moreover, in (\ref{eq:csv_auxvar}) we have also introduced the source term $\bS$. In this work $\bS$ contains a forcing term $\mathbf{f}$ which is added to preserve the correct mass flux along the channel, its expression is given by 
\[\bS = \left[\begin{array}{c}
0 \\
 \lrho\hyb \mathbf{f} \\
 \gamma Ma^{2 } \lrho\hyb \mathbf{f} \cdot \langle\wt{\bu}\hyb
 \end{array}
 \right]
 \] .

Integrating (\ref{eq:csv_auxvar}) and multiplying by the test functions $v$ and $\br$, we obtain the  weak form
\begin{subequations}\label{eq:csv_auxvar_wf}
\begin{align}
\int_{\Omega} v \de_t \bU d\Omega & - \int_{\Omega} \bF(\bU,\bmG)\cdot\nabla v d\Omega + \int_{\partial \Omega} \bF(\bU,\bmG) \cdot \bn v d \sigma  = \int_{\Omega}\bS d\Omega\\[2mm]
& \int_{\Omega} \br \bmG d\Omega+ \int_{\Omega} \boldsymbol \varphi \nabla\cdot{\br} d\Omega  - \int_{\partial \Omega} \boldsymbol \varphi \bn \cdot \br  d \sigma = 0,
\end{align}
\end{subequations}
where the fluxes $\bF^{{\rm c}}$,$\bF^{{\rm v}}$ and $\bF^{{\rm sgs}}$ are collected in $\bF=\bF^{{\rm c}}-\bF^{{\rm v}}-\bF^{{\rm sgs}}$. 

For the discretization we follow  the method of lines: we start from space discretization and then we use a time integrator to advance in time. In this case a Strongly Stability Preserving Runge–Kutta method (SSPRK) ~\cite{spiteri:2002} has been used.

As usual, to obtain the DG discretization , we consider a tessellation  $\mT_h$ of the computational domain $\Omega$ into non-overlapping tetrahedral elements $K$. We also introduce the finite element space of the polynomial functions of degree at most $q$ on the element $K$, which is defined as
\begin{equation}\label{eqn:mV_def}
\mV_h = \left\{ v_h \in L^2(\Omega): v_h|_K \in \mathbb{P}^q(K), \,
\forall K\in\mT_h \right\}.
\end{equation}
Therefore, the DG formulation for problem (\ref{eq:csv_auxvar_wf}) will be:  find the solution $(\bU_h,\bmG_h)\in(\,(\mV_h)^5\,,\,(\mV_h)^{4\times3}\,)$ such that,
$\forall K\in\mT_h$, $\forall v_h\in\mV_h$, $\forall
\br_h\in(\mV_h)^3$,

\begin{subequations}
\label{eq:DG-space-discretized}
\begin{align}
\displaystyle
\frac{d}{dt}\int_K \bU_h v_h\,d\bx
& \displaystyle
- \int_K \bF(\bU_h,\bmG_h)\cdot\nabla v_h\, d\bx
\nonumber \\[3mm]
& \displaystyle
+ \int_{\partial K} \widehat{\bF}(\bU_h,\bmG_h)\cdot \bn_{\partial K} v_h\,
d\sigma
= \int_K \bS v_h \,d\bx,
\\[3mm] \displaystyle
\int_K \bmG_h \cdot \br_h \,d\bx
& \displaystyle
+ \int_K {\boldsymbol \varphi_h}\nabla\cdot\br_h\, d\bx
\nonumber \\[3mm]
& \displaystyle
- \int_{\partial K} \widehat{\boldsymbol \varphi} \bn_{\partial
K}\cdot\br_h \, d\sigma = 0,
\end{align}
\end{subequations}
where $\bU_h=[ \rho_h\,,\rho_h\bu_h\,,\rho_he_h ]^{T}$,$ \boldsymbol \varphi_h = [\bU_h, T_h]^{T}$, $\bn_{\partial K}$ represents the outward normal on $\partial K$ and the terms $\widehat{\bF}$ and $\widehat{\boldsymbol \varphi}$ are the numerical fluxes. These terms represent the only connection between adjacent elements, which would be otherwise uncoupled. The numerical fluxes are needed to solve the ambiguity of double valued functions at the interface between adjacent elements and to weakly impose the boundary conditions on $\partial \Omega$.
There are different ways to define the numerical fluxes ~\cite{giraldo:2008}, in this work we use the \emph{Rusanov} flux for $\widehat{\bF}$ and the centered flux for $\widehat{\boldsymbol \varphi}$.

The solution and the test functions are defined in terms of orthogonal basis functions, this is a quite natural choice considering that in DG there are no constrains related to the continuity; this approach is commonly defined as modal DG. We also mention that all the integrals are evaluated by means of the  quadrature formulae  reported in ~\cite{cools:2003}. In order to have a correct evaluation for the products, we have used formulae which are exact for polynomial of degree up to $2q$.

The unknowns in (\ref{eq:fnscompr}) are  filtered quantities, in particular the Favre average defined in (\ref{eq:favre}) has been used.  Nevertheless, according to the concept of implicit filtering previously mentioned, no explicit hybrid  filter is applied. Therefore, the unknowns are directly computed as $\langle\tilde{ \cdot} \hyb$. Regarding the LES modelling,a common strategy is to associate the filter size to the grid resolution, including the filtering process into the spatial discretization. Using a DG formulation, this approach can be extended considering the degree $q$ of polynomial basis functions used to define the solution in each element. By doing this, it is possible to  enlarge the filter size projecting the solution on basis function of lower degree and this operation becomes trivial using orthogonal basis functions: in fact, it is obtained simply zeroing the last coefficients of the local expansion. This approach is very useful for the implementation of the dynamic procedure ~\cite{germano:1991} in which two different levels of filter are required. These guidelines have been followed for pure LES (for a detailed description we refer to \cite{abba:2014})  and also to determine LES subgrid terms in the hybrid model, but in this case the LES model coefficients have been computed from $\langle\tilde{ \cdot} \hyb$ variables instead of $\langle\tilde{ \cdot} \les$. 

\section{Results and discussion}

\begin{table}
\centering
\footnotesize{
\begin{tabular}[hbtp]{|l|c|c|c|}
\hline
& Moser  et Al&  Present & Present \\
& (MKM) & \emph{coarse} & \emph{fine} \\
 \hline
$\MA_{{\rm b}}$ & \textemdash & 0.2 & 0.2 \\
 \hline
$\RE_{{\rm b}}$ & 2800 & 2800 & 2800   \\
 \hline
 $L_x$ & $4\pi$ &  $2\pi$ & $2\pi$  \\
 \hline
 $L_z$ & $\frac{4}{3}\pi$ & $\frac{4}{3}\pi$& $\frac{4}{3}\pi$  \\
 \hline
 $\Delta_x^+$ & 17.7 &  23 & 18.4  \\
 \hline
 $\Delta_z+$ & 5.9 & 10 &  8.57 \\
 \hline
 $\Delta^+_{y_{min}}/\Delta^+_{y_{max}}$ &  0.05/4.4 & 0.65/7.9 & 0.65/5.20 \\
\hline
\end{tabular}
}
\caption{Parameters and grid characteristic for simulations and reference test case}
\label{tab:parameter}
\end{table}

The test case considered for the simulations is the turbulent channel flow at Ma =$0.2$ and the numerical results  were compared to LES and DNS data. The latter has been obtained by the incompressible numerical simulation of Moser et al. (MKM) \cite{moser:1999}.

Two different values of blending factor $k$ for the hybrid method have been tested: $k=0.5$ and $k=0.75$. As previously mentioned,  in both cases the anisotropic dynamic model \cite{abba:2003} has been used as LES model. The same model has been employed in pure LES computation.

The simulations herein performed are realized using the finite element toolkit {\tt FEMilaro} \cite{femilaro}, a FORTRAN/MPI library, available under GPL license. 
 
The computational domain size, in dimentionless units, is $2\pi \times 2\times 4/3\pi$, representing respectively $L_x$, $L_y$ and $L_z$.We use $x$ for streamwise direction, $y$ for normal direction and $z$ for spanwise direction. The bulk Reynolds number, computed with the half height of the channel, is $\RE_{{\rm b}} = \frac{\rho_{{\rm b}} U_{{\rm b}} d}{\mu_{{\rm w}}}=2800$. No-slip, isothermal boundary conditions have been prescribed at the wall, $ y=\pm1$, while periodic conditions have been applied for the remaining directions. 

Two different grids have been used, the first grid, named \emph{coarse}, has $N_x=8$, $N_y=16$, $N_z=12$ hexahedra in the $x, y, z$ directions, while for the second grid, named \emph{fine}, we have $N_x=10$, $N_y=24$, $N_z=14$. Each hexahedra is divided into $N_t=6$ tetrahedral elements which form the structured mesh. These two meshes are uniform in $x$ and $z$ directions, while, to increase the resolution near the wall, in the normal direction ($y$) the planes that define the hexahedra  are given by:

\begin{equation}\label{eqn:stretch}
 y_j = -\dfrac{\tanh\left(\w\left(1-2j/N_y\right)\right)}
              {\tanh\left(\w\right)} \qquad  j=0, \ldots, N_y,
\end{equation}
where the parameter $\w$ is set fixing the position of the first element.

Mesh resolution can be estimated using the following formula:

\begin{equation}
\label{eq:delta}
\Delta_i=\frac{H_i}{\sqrt[3]{N_t N_q}} \quad \quad i= x,y,z ,
\end{equation}
where $H_i$ represents a characteristic element size and $N_q$ is the number of degrees of freedom for each finite element, in this case employing $4^{th}$ degree basis functions we have $N_q=35$. Multiplying (\ref{eq:delta}) by $Re_{\tau}$, i.e. the skin friction Reynolds number ($Re_{\tau}\approx 180$ for the simulations performed), we obtain the grid spacing estimation  in wall unit, $\Delta^{+}_{i}$, reported in Table ~\ref{tab:parameter}.

All the considered numerical simulations start from a laminar Poiseille profile. The turbulence is obtained adding a perturbation to the velocity in the $x$  direction. This random perturbation is computed from a fixed number of iteration of logistic map: $\xi^{k+1}=3.999 \xi^{(k)}(1 - \xi^{(k)} )$. As result, we can obtain a definition of the random perturbation which allows the repeatability of the results. After the statistical steady turbulent regime was reached, the simulations were continued enough to have a well verified time invariance for the mean profiles. In the simulations herein shown the sample used for statistics computation is $60$ non-dimentional time units.  

The statistics are computed averaging the solution, both in space and time, on a set of fixed planes, parallel to the wall. For a generic quantity $\varphi$ we have:

\begin{equation}
<\varphi>(|y|) = \frac{1}{2TL_xL_z} \int_{t_f}^{t_f -T} \int_0^{L_x}
\int_0^{L_z} \left(\varphi(t,x,-|y|,z) +
\varphi(t,x,|y|,z)\right)\,dz\,dx\,dt.
\label{eq:channel-average}
\end{equation}
where $T$ is the time used for statistics computation.

In order to maintain a constant mass flux along the channel a body force in streamwise direction has been added. This forcing term $f_x(t)$ is proportional to the difference between the mass flux calculated at each time step $Q(t)$ and the prescribed value $Q_0$:

\begin{equation}\label{forcing}
f_x(t) = - \frac{1}{\rho_{{\rm b}}} \left[ \alpha_1 \left( Q(t) - Q_0 \right) + \alpha_2 \int_0^t
\left( Q(s)- Q_0 \right) ds \right],
\end{equation}
the constants $\alpha_1$ and $\alpha_2$ are respectively $0.1$ and $0.2$
  
\begin{figure}
\includegraphics[width=.32\textwidth]{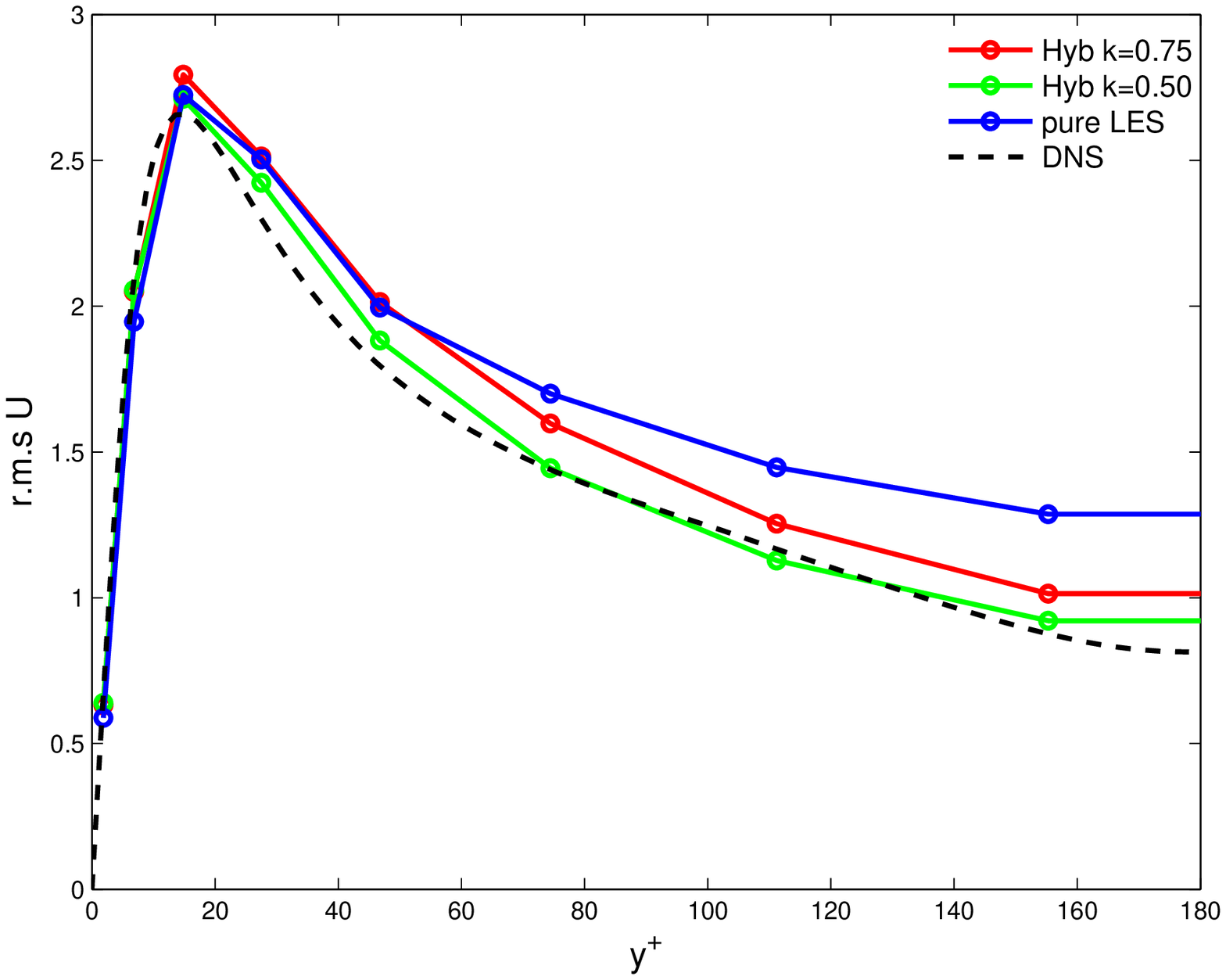}
\includegraphics[width=.32\textwidth]{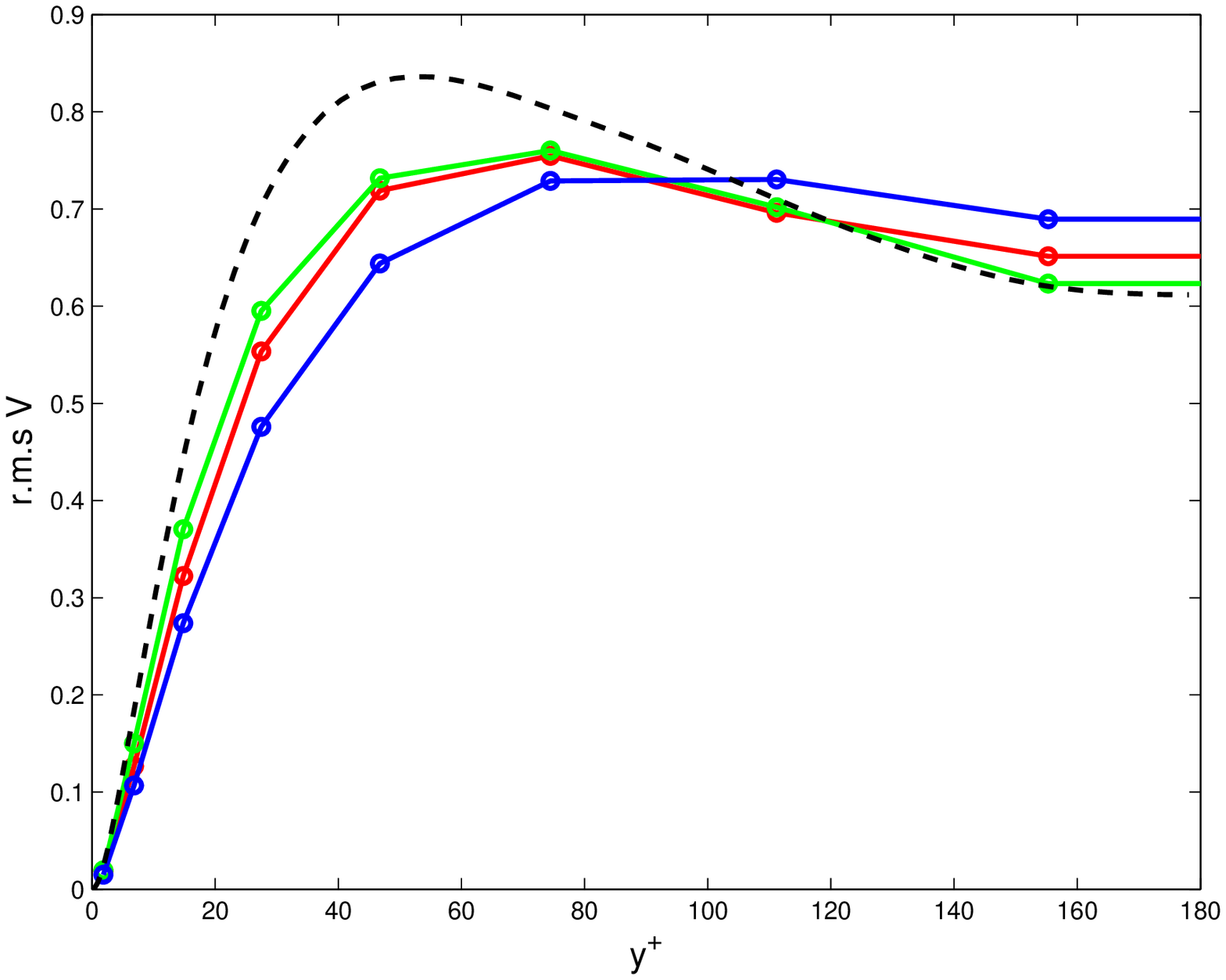}
\includegraphics[width=.32\textwidth]{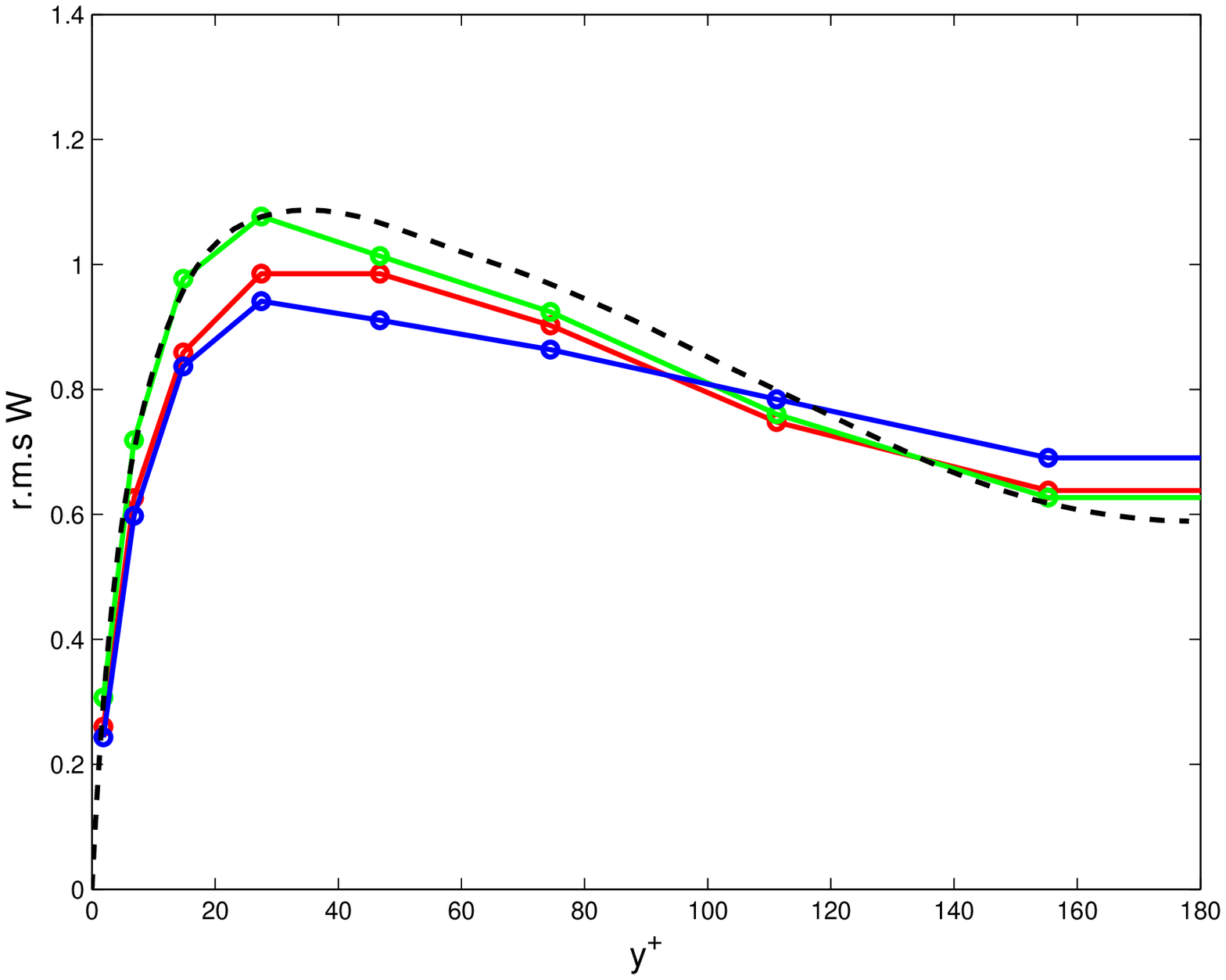}
\caption{Velocity \emph{r.m.s} profiles, coarse grid: \emph{left} streamwise component,\emph{center} normal component, \emph{right} spanwise component}
\label{fig:rms_std}
\end{figure}

\begin{figure}
\includegraphics[width=.32\textwidth]{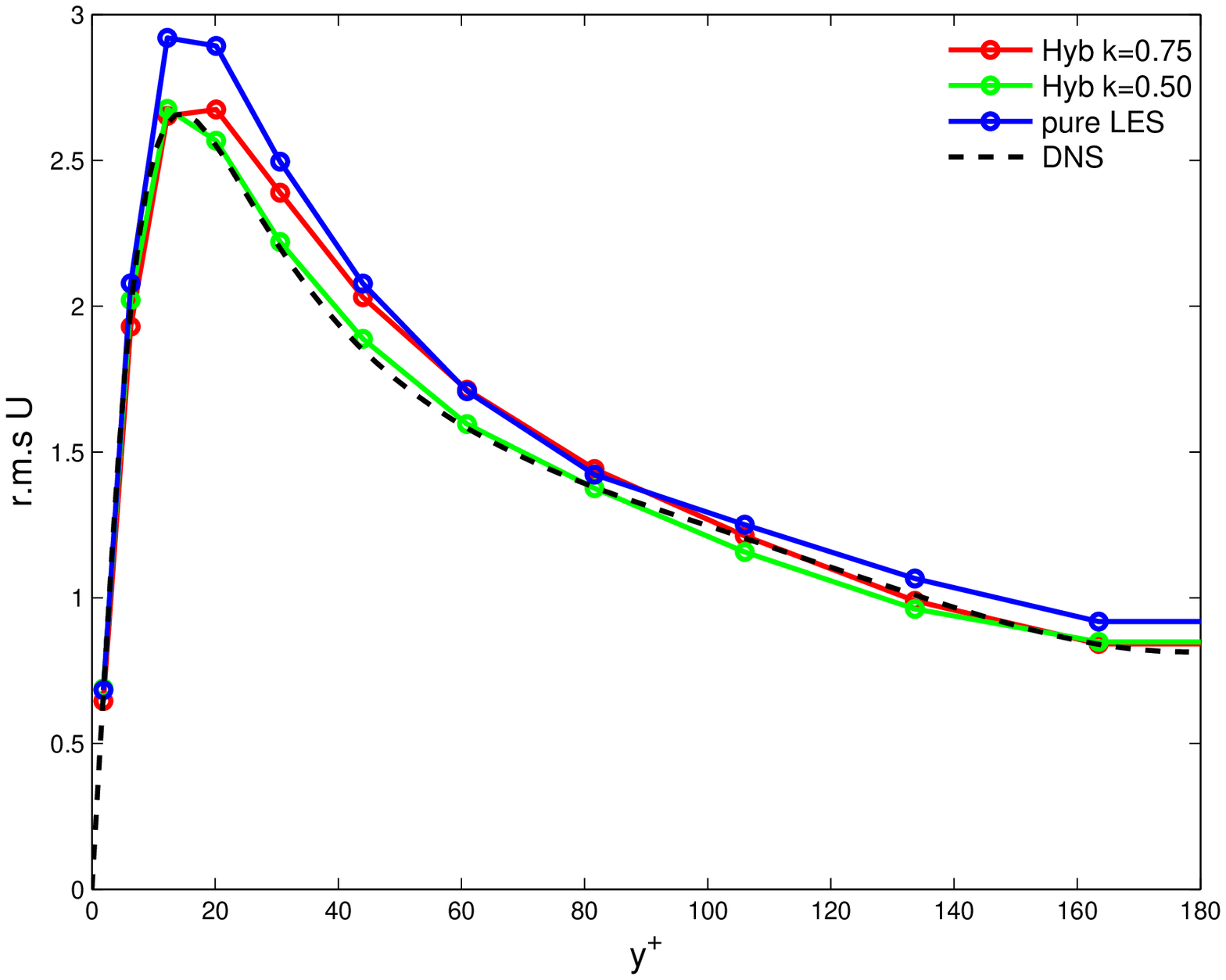}
\includegraphics[width=.32\textwidth]{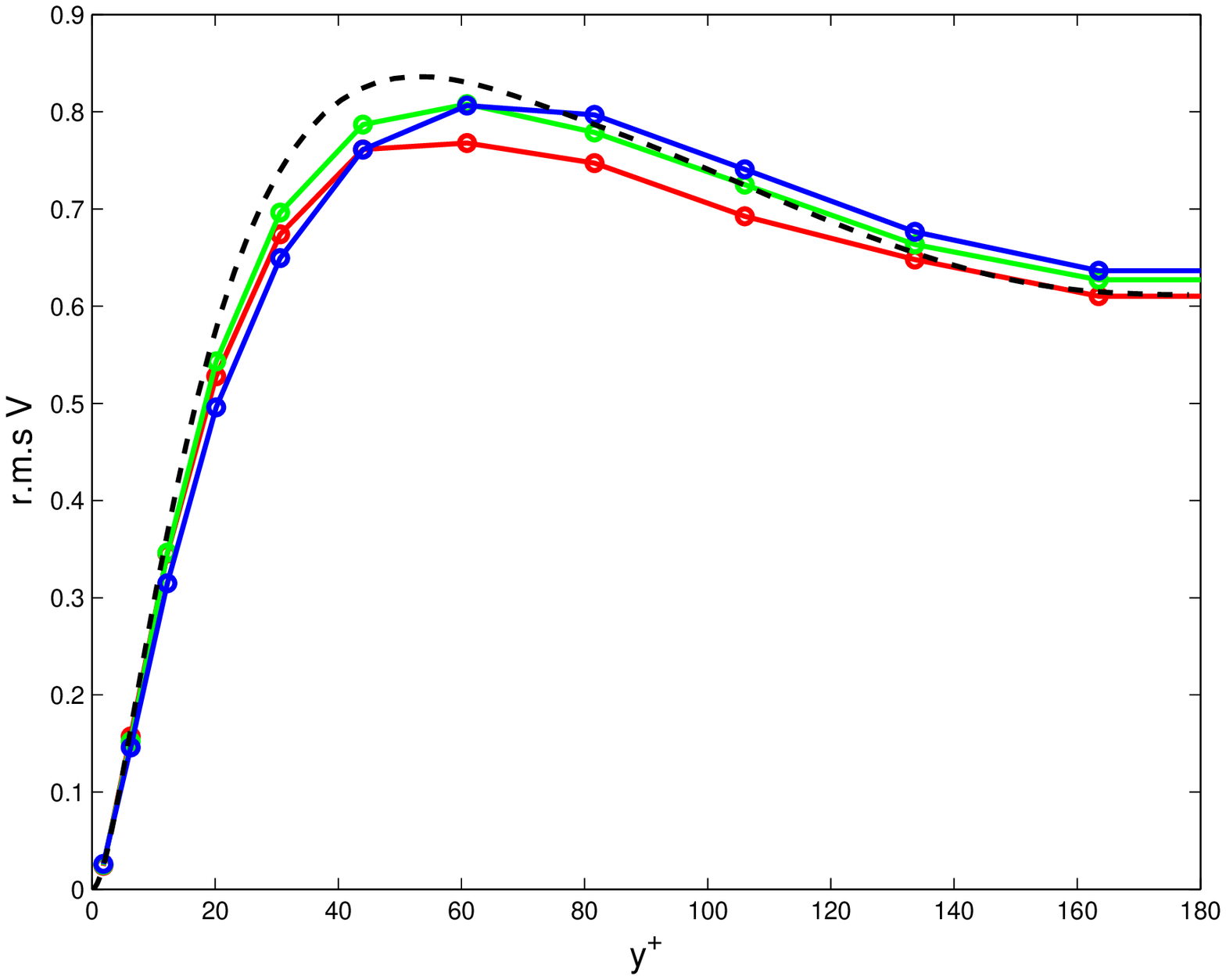}
\includegraphics[width=.32\textwidth]{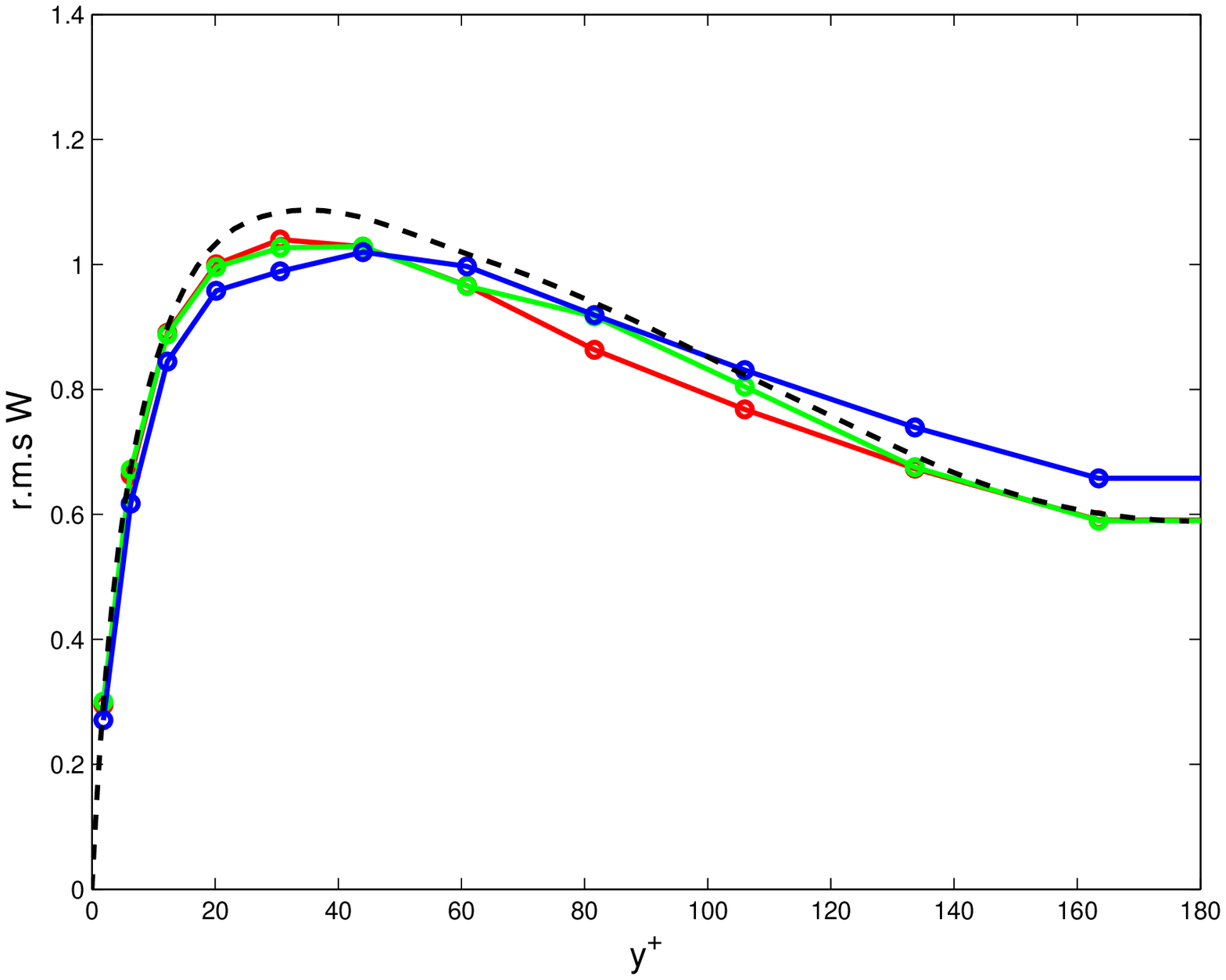}
\caption{Velocity \emph{r.m.s} profiles, fine grid: \emph{left} streamwise component,\emph{center} normal component, \emph{right} spanwise component}
\label{fig:rms_fine}
\end{figure}

Figures ~\ref{fig:rms_std} and ~\ref{fig:rms_fine} show the \emph{root mean square} for the velocity in $x$,$y$ and $z$ directions. The results of the hybrid method appear to be better than those obtained with the pure LES. In particular, the simulations with $k=50$ are in very good agreement with DNS data also for the \emph{coarse} grid. As expected, the results of the \emph{fine} grid are closer to DNS and the differences between pure LES, $k=50$ and $k=75$ are reduced. The only exception is represented by the peak in Fig. \ref{fig:rms_fine} for LES and in $k=0.75$ in the $u$ profile at $y^{+}\approx 20$. Probably, one of the reasons for this behaviour can be related to the greater anisotropy of the fine grid: in fact, the increase of resolution in the $x$ direction is significantly lower then the one in the $y$ direction. 

\begin{figure}
\includegraphics[width=.46\textwidth]{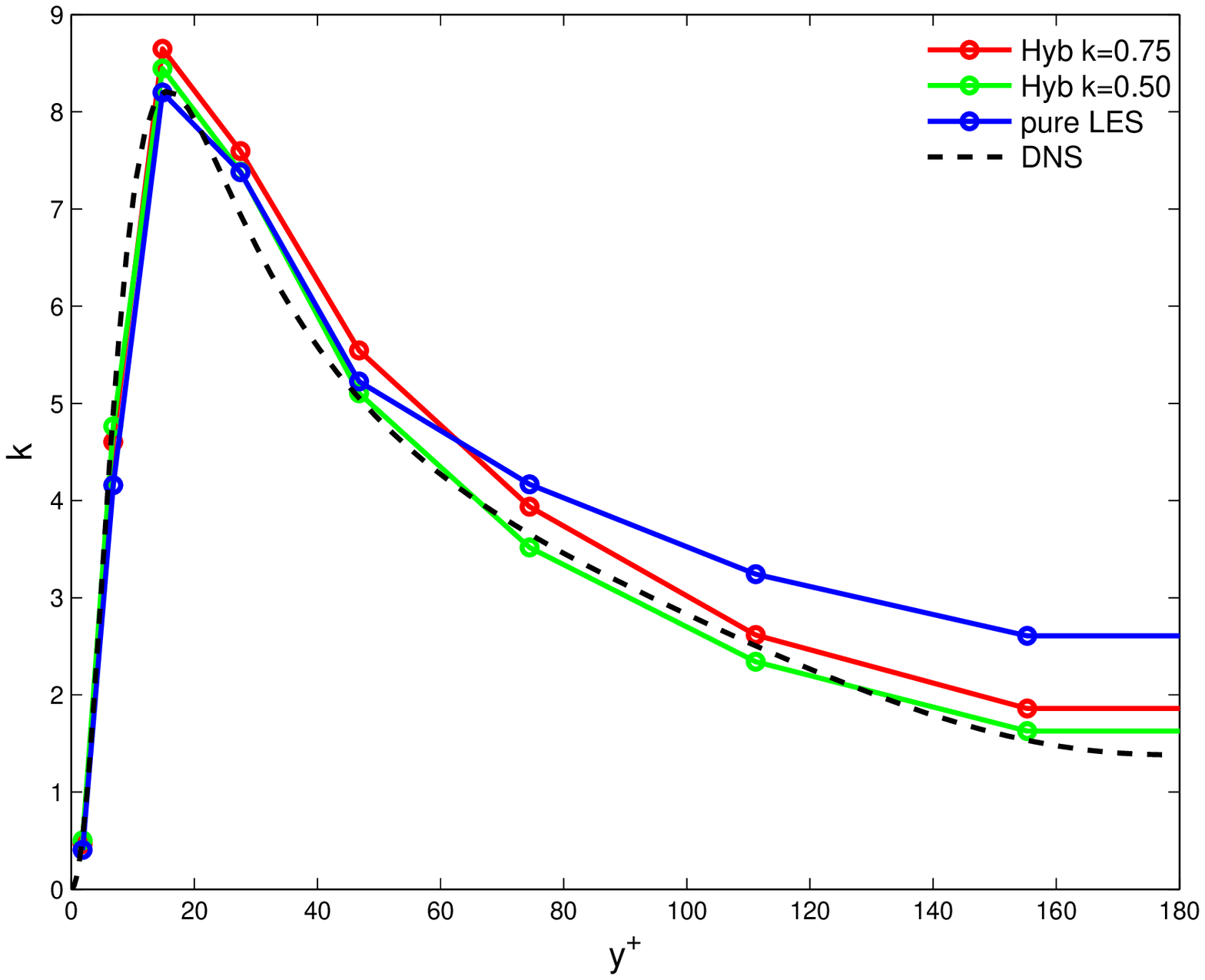}
\includegraphics[width=.46\textwidth]{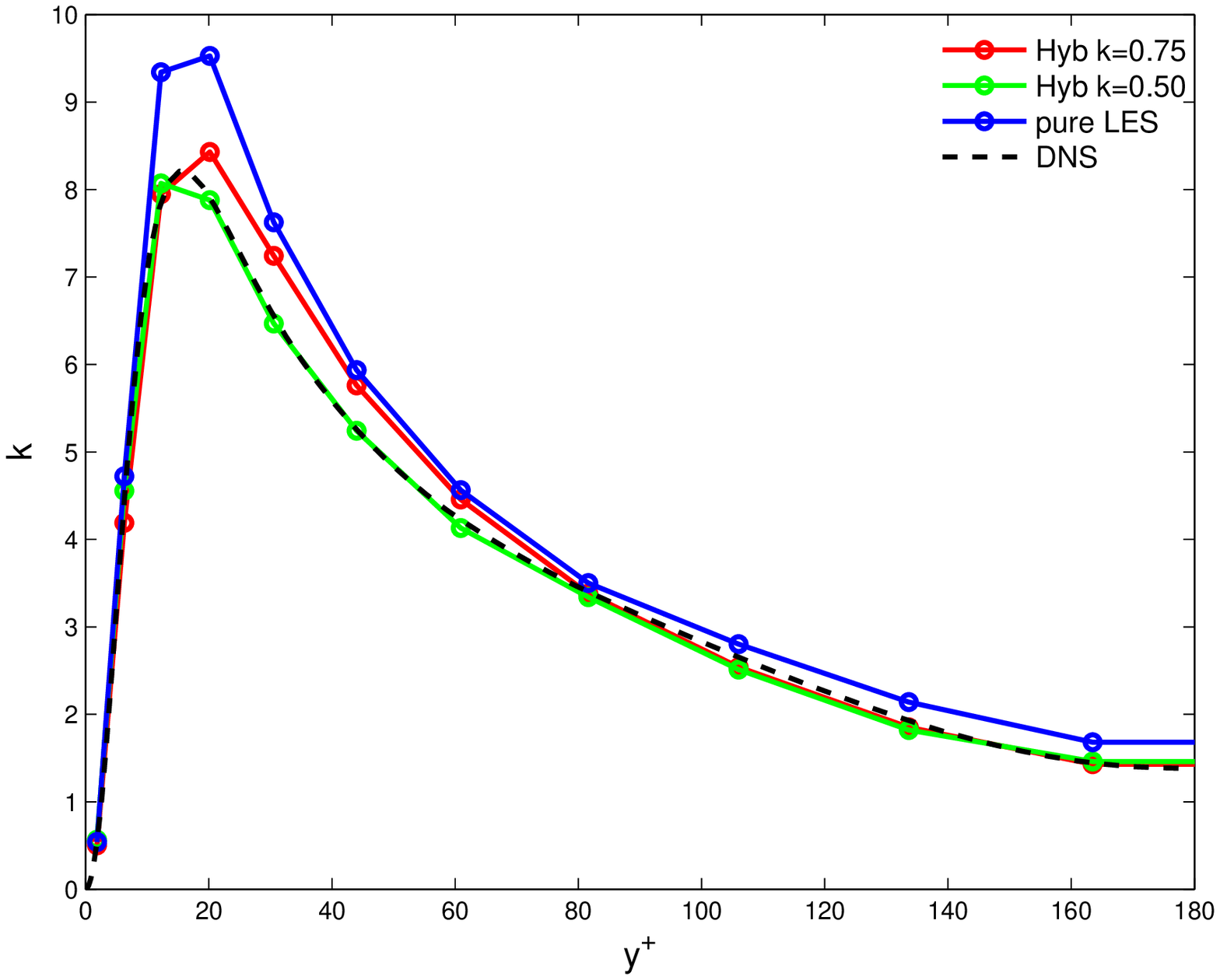}
\caption{Turbulent kinetic energy profiles, $k$: \emph{left} coarse grid, \emph{right} fine grid}
\label{fig:tke}
\end{figure}

The turbulent kinetic energy profiles (Fig. \ref{fig:tke}) are strongly dependent on the streamwise velocity component, so the results are similar to the $r.m.s.$  profiles seen before. Regarding the \emph{fine} grid, beyond $y^{+} =80$, the results of the hybrid methods are almost identical and in good agreement with the DNS. Closer to the wall, the $k=0.75$ and LES profiles get worse and show the same peak previously mentioned. Notice that, also for turbulent kinetic energy, the results for $k=0.50$ are very close to DNS on both grids.

\begin{figure}
\includegraphics[width=.46\textwidth]{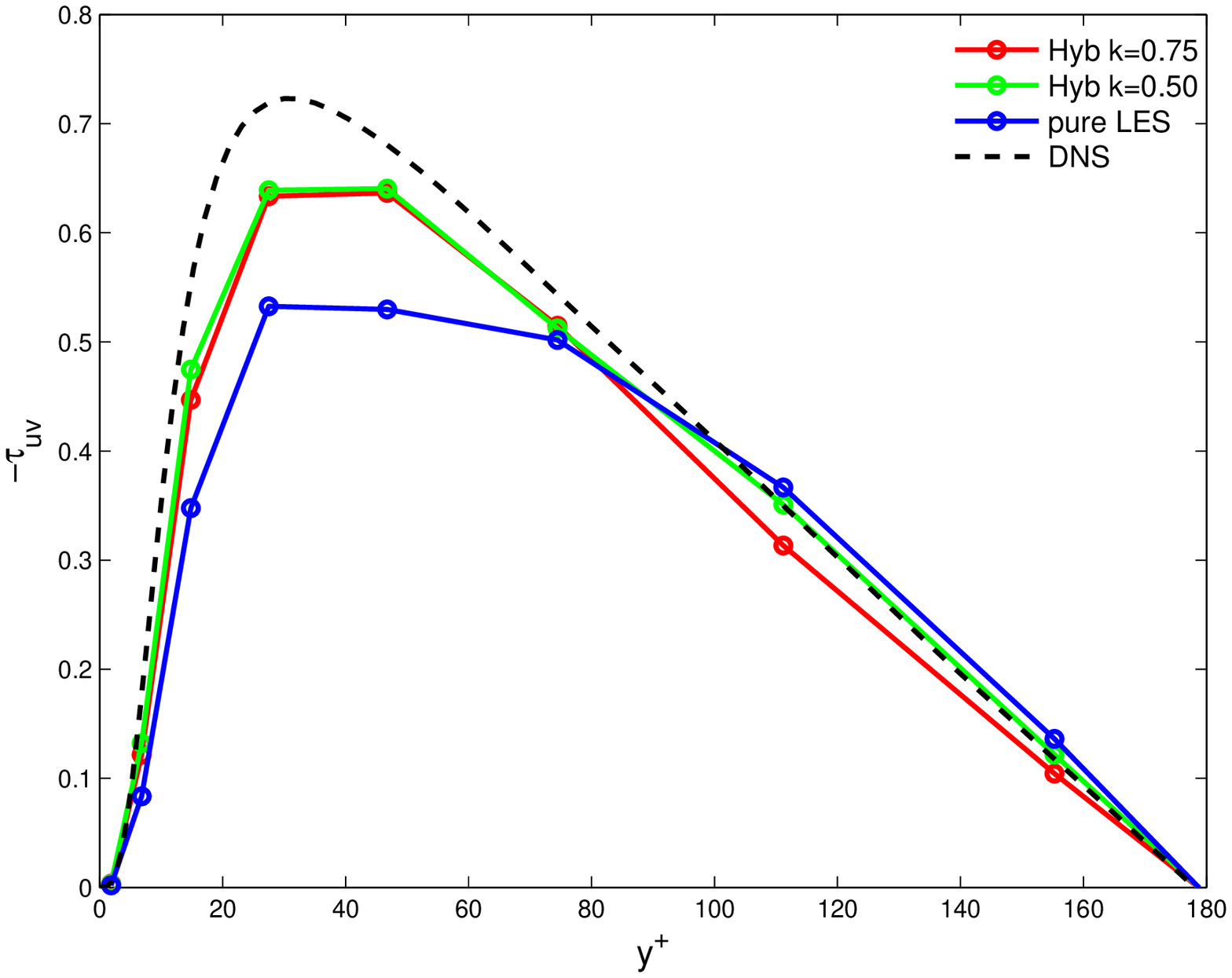}
\includegraphics[width=.46\textwidth]{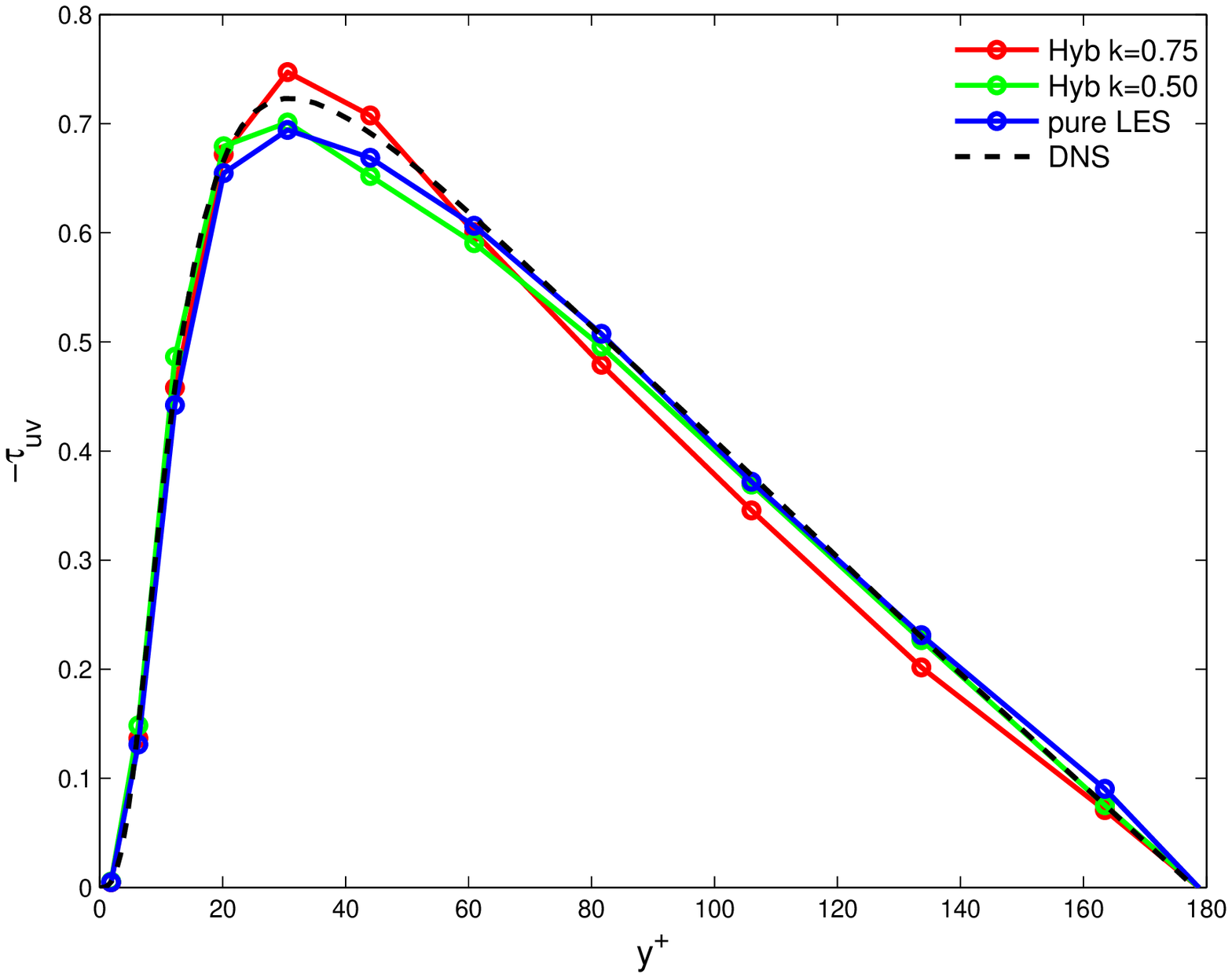}
\caption{Shear stress profiles, $\tau_{uv}$: \emph{left} coarse grid, \emph{right} fine grid}
\label{fig:tau_uv}
\end{figure}

Fig. \ref{fig:tau_uv} shows the shear stress profile. In this case the results obtained using hybrid method in the \emph{coarse} grid are significantly better then LES. In the \emph{fine} grid the three simulations give similar results and are in good agreement with DNS.

\begin{figure}
\includegraphics[width=.46\textwidth]{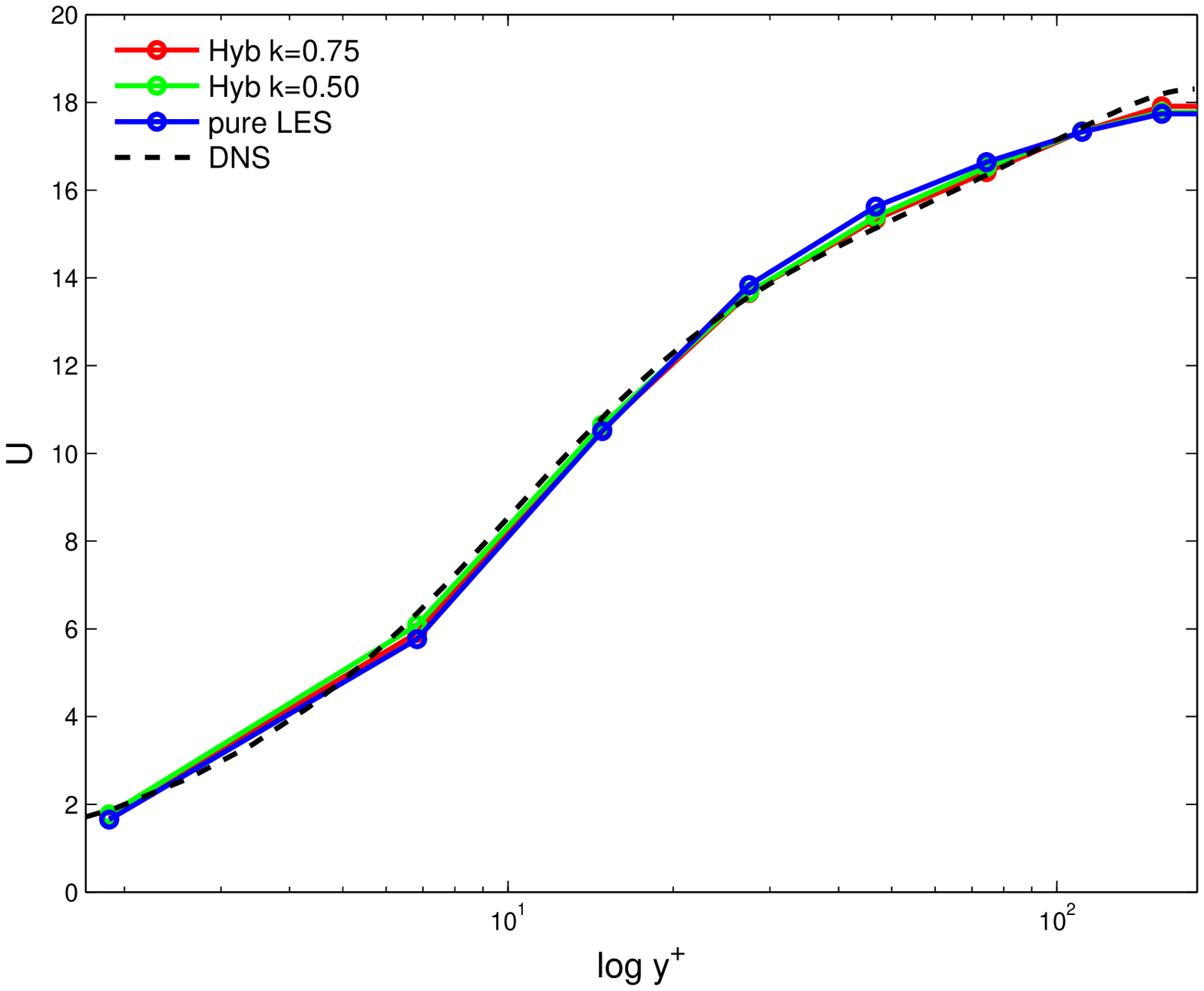}
\includegraphics[width=.46\textwidth]{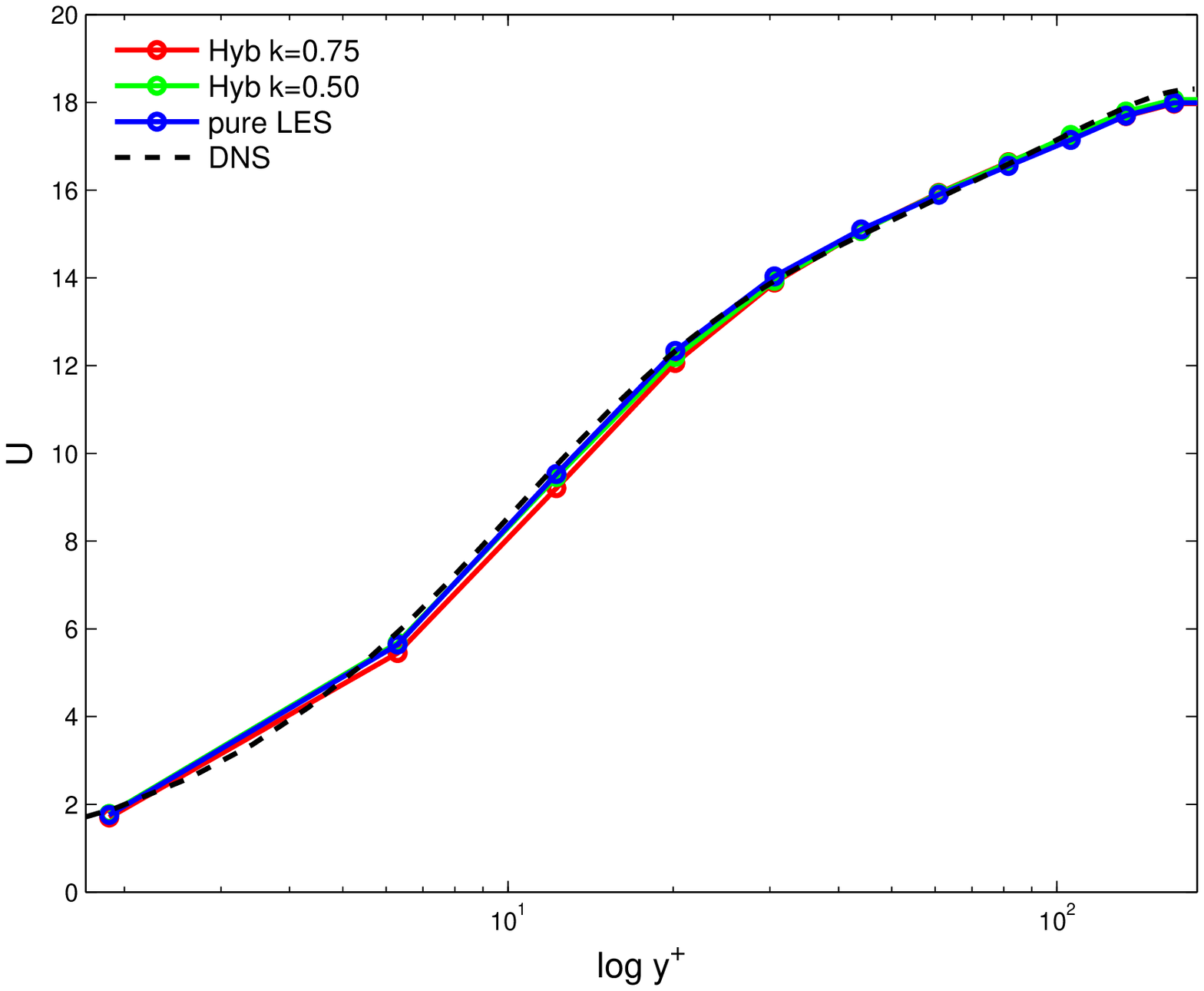}
\caption{Velocity profiles, semi--logarithmic scale: \emph{left} coarse grid, \emph{right} fine grid}
\label{fig:u_log}
\end{figure}

Finally, in  Fig. \ref{fig:u_log} velocity profiles are shown. The semi--logarithmic scale does not show remarkable differences between the cases studied. For the coarse grid, and partially for the fine grid too, we have an underestimation of the velocity at the centerline. 

The results highlight a general improvement obtained with the hybrid method with respect to pure LES. This points out that the additional reconstructed RANS term can be suitable to integrate the LES results. In fact, for the coarse grid simulation, where we have a smaller quantity of resolved energy,  the improvement obtained with the introduction of hybrid terms is greater. Moreover, to confirm this, the better results have been obtained with the lower $k$, i.e. where the hybrid terms, and then also the RANS term, are more important.

In our opinion, an interesting point is that, different from what we would expect, $k=0.75$  are not in general closer to LES then $k=0.50$. This shown the complexity of the interaction between LES and RANS, we plan to further investigate this issue.

\section{Concluding remarks}
We have studied and tested a RANS reconstruction technique for Germano's hybrid filter approach. 
Tests have been conducted for the turbulent channel at Ma $=0.2$, considering two different constant blending factors: $k=0.75$ and $k=0.50$; and two computational grids.
The RANS/LES method has been implemented using a variational multiscale approach combined to a DG-FEM space discretization.

The results obtained with the hybrid method are quite promising. In fact,  they show a better agreement with the DNS results compared to the LES computations, especially for the coarser grid. Therefore, this preliminary work shows that the hybrid RANS reconstructed  model can be suitable for turbulence description. Moreover, it confirms the potentiality of the DG-FEM approach for fluid dynamics and more specifically for LES. 

Future works will be focused on a space-depending blending factor, this will lead to several extra terms related to the non-commutativity between the hybrid filter and the spatial derivatives. We plan also to perform a comparison between hybrid methods with RANS reconstruction and hybrid methods coupled with an explicit RANS method, in order to better analyse the  benefits and drawbacks of the procedure herein proposed, and to extend this work to more compressible flows.

\section*{Acknowledgments}
The numerical results shown in this paper has been obtained with the computational resources provided  by CINECA (Italy) and NIIF(Hungary), respectively within the high performance computing projects ISCRA-C LES-DiG and DECI-11 HyDiG.

\bibliographystyle{unsrt}

\bibliography{DG_hybrid}




\end{document}